\documentclass[aps,prb,twocolumn,showpacs]{revtex4}
\usepackage{graphicx}

\begin{document}

\title{Vortex-unbinding and finite-size effects in
Tl$_2$Ba$_2$CaCu$_2$O$_8$ thin films}

\author{Hao Jin}
\email{hjin@ssc.iphy.ac.cn}

\author{Hai-Hu Wen}
\email{hhwen@aphy.iphy.ac.cn}

\affiliation{National Laboratory for Superconductivity, Institute of
Physics, Chinese Academy of Sciences, P.~O.~Box 603, Beijing
100080, P.~R.~China}

\date{\today}

\begin{abstract}
Current-voltage ($I$-$V$) characteristics of Tl$_2$Ba$_2$CaCu$_2$O$_8$
thin films in zero magnetic field are measured and analyzed with the
conventional Kosterlitz-Thouless-Berezinskii (KTB) approach, dynamic
scaling approach and finite-size scaling approach, respectively. It is found
from these results that the $I$-$V$ relation is determined by the vortex-unbinding
mechanism with the KTB dynamic critical exponent $z=2$. On the other hand, the
evidence of finite-size effect is also found, which blurs the feature of a phase transition.
\end{abstract}


\maketitle

\section{Introduction}
For two-dimensional (2D) systems in which the conventional long-range order is
absent \cite{Mermin}, the definition of topological order and Kosterlitz-Thouless-Berezinskii
(KTB) transition \cite{Berezinskii,Kosterlitz,Kosterlitz1} has been proposed,
with the latter characterized by a sudden change in the response of the system to
the external perturbation. This type of transition was originally not considered to occur in
superconductors \cite{Kosterlitz} due to the dimensional restriction from the 2D penetration depth
$\lambda_\perp$($\simeq \lambda^2/d$, with $\lambda$ the bulk penetration depth and $d$ the
thickness of the film \cite{Pearl}), beyond which the logarithmic intra-pair binding interactions
of vortices are saturated and free vortices will appear below the vortex-unbinding temperature.
However, it was later argued that the KTB transition occurs in superconductors \cite{Halperin}
in a practical sense if $\lambda_\perp$ exceeds the system size, and this has been supported by
experiments in dirty conventional superconductors \cite{Hebard} with the enhanced $\lambda_\perp$.

For cuprate superconductors characterized by layered structure and high anisotropy, there have
been early theoretical works about the KTB issue \cite{Rasolt}. 
And the phase fluctuation has been argued to play an important role due
to the low superfluid density \cite{Emery}. This is supported by experiments including the
high-frequency conductivity in underdoped Bi$_2$Sr$_2$CaCu$_2$O$_8$ (BSCCO)
\cite{Corson}, which revealed that the properties are dominated by the vortex-unbinding in the
samples and that the short-time phase correlations persist above the superconducting transition temperature
$T_{\rm c}$ and even up to the pseudogap temperature $T^*$, and the Nernst effect in underdoped
La$_{2-x}$Sr$_x$CuO$_4$ and Bi$_2$Sr$_{2-y}$La$_y$CuO$_6$ \cite{Xu,Wang} reveals the
existence of vortex-like excitation till the resolution limit (at temperatures 50-100 K above $T_{\rm c}$).
The experiments probing the phase coherence have been considered much helpful in the research of
KTB transition and high-$T_{\rm c}$ superconducting mechanism \cite{Millis}.

As to the zero-field electrical transport characteristics, KTB theories have derived that \cite{Minnhagen,Halperin},
in the zero current ($I \rightarrow 0$) limit, the exponent $a$ of current-voltage ($I$-$V$)
relation, $V=I^a$, will jump from 3 to 1 at the KTB transition temperature $T_{\rm KT}$, and the
linear resistance $R_{\rm lin}$($\equiv \lim_{I \rightarrow0}V/I$) is proportional to
$\exp[-2\sqrt{b/(T/T_{\rm KT}-1)}]$, where $b$ is a non-universal constant. These features in the
thermodynamic limit have been originally used in the determination of KTB transition, and is usually
referred to as `conventional approach'.

Besides, the dynamic scaling approach proposed by Fisher, Fisher and Huse \cite{Fisher} (FFH) has
also been used in the study of KTB transition. In this ansatz, the data at finite currents are used in the
scaling function, in which the dynamic critical exponent $z$, characterizing the relation between the
relaxation time $\tau$ and the correlation length $\xi$ as $\tau=\xi^z$, equals 2 for a 2D superconductor
in zero magnetic field. With the connection $a=1+z$ at the KTB transition temperature, this corresponds
to $a=3$ in the conventional approach.

The experimental transport results of KTB transition in cuprates remain controversial. There have been
reports of KTB transition in YBa$_2$Cu$_3$O$_{7-\delta}$ (YBCO) \cite{Matsuda}, BSCCO \cite{Martin},
and Tl$_2$Ba$_2$CaCu$_2$O$_8$ (TBCCO)\cite{Kim,Wen} systems. However,
nanovolt-level current-voltage ($I$-$V$) measurements by Repaci {\it et al.}\cite{Repaci} on single-unit-cell
YBCO films show ohmic behavior far below the nominal KTB temperature. There are even results of
$z \approx 6$, as given by Ammirata {\it et al.} \cite{Ammirata} for the transport measurements
of thin BSCCO crystal.

Holzer {\it et al.} \cite{Holzer} simulated $I$-$V$ curves for 2D Josephson junction arrays, and
argued that for the finite-size effect data, the scaling parameters $z$ and $T_{\rm KT}$ depends
critically on the noise floor of the measuring system, and that the value $z \approx 6$
for a variety of 2D systems is related to the nature of data collection or the resolution of
instrument, rather than characterizing some universality of physics.

Medvedyeva {\it et al.} \cite{Medvedyeva} argued that the FFH dynamic scaling
method is only appropriate for those phase transition with a finite correlation
length $\xi$, thus is not compatible with a system undergoing a KTB
transition with $\xi=\infty$ at temperatures below $T_{\rm KT}$.
Instead they took the finite-size effect into account and argued that,
for a 2D system with linear size $L$, the scaling behavior should be
dependent on $L$ rather than $\xi$ at low temperatures for which
$L<\xi(T)$, hence the electric field $E$ and the applied current density
$J$ can be scaled as $E/JR_{\rm lin}=h(JLg_L(T))$ for small values of $J$,
where $h(x)$ is the scaling function and $g_L(T)$ is the function that
makes the $E/JR_{\rm lin}$-$JLg_L(T)$ curves collapse. Moreover, they performed
the numerical simulations on 2D finite size ($L \times L$) systems with
resistive-shunted-junction (RSJ) model, and re-analyzed the experimental
data of Repaci {\it et al.}. Both numerical and experimental results suggest $z=2$.
At the same time, an exponent $\alpha_L(T)$, which may depend on the system
size $L$ and the temperature $T$, has been defined in the finite-size scaling scenario
by $g_L(T)\equiv A_LR_{\rm lin}(T)^{-\alpha_L(T)}$, where $A_L$ is a universal
constant for a given system with size $L$. Then it comes out that if $\alpha_L(T)$
is a constant ($\alpha_L(T)=\alpha$) for all the available data over a certain
temperature region, the resistance will seem to vanish at the
temperature for which $z(T)=1/ \alpha$. For both 2D RSJ model and Repaci
{\it et al.}'s data, it was found that $\alpha \approx1/6$ in a limited temperature
region  by using the finite-size scaling approach, which amounts to $z \approx 6$.
However, on the analogy of the results
of 2D RSJ model, it was concluded that there is always a linear resistance at
any temperatures for a finite-size system, and $\alpha \approx1/6$ is only satisfied
in a limited range, so the result $z \approx6$ should not be considered as the symbol
of a real transition.

In view of these controversies, we have carried out the research on
Tl$_2$Ba$_2$CaCu$_2$O$_8$ thin films. The $R$-$T$ and $I$-$V$
characteristics was measured in zero magnetic field. This paper will begin
the analyses with the conventional approach which focuses on the
slope $a$ of the logarithmic $I$-$V$ curves in the $I \rightarrow 0$
limit, then the FFH dynamic scaling method will be used in analyzing
the experimental data. Since the dynamic scaling method is argued to
have some flexibility, the criteria of a real KTB transition proposed
by Strachan {\it et al.}\cite{Strachan}~have been used. Finally, we will
follow the finite-size scaling procedures of Medvedyeva {\it et al.}. It is
found that the experimental results are consistent with not only
thermodynamic limit for which $z=2$, but also the exponent
$\alpha \approx 1/2$, where $\alpha$ is the exponent defined by
Medvedyeva {\it et al.}. On the other hand, the evidence of the finite-size
effect is also found, suggesting a crossover rather than a real phase
transition  near the vortex-unbinding temperature $T_{\rm KT}$.

\section{Experiments}
The Tl$_2$Ba$_2$CaCu$_2$O$_8$ films were prepared by a two-step
procedure on (001) LaAlO$_3$ substrates. Details on the
fabrication of the films have been published previously \cite{Yan}.
X-ray diffraction patterns (XRD) taken from the samples show that
only (00$l$) peaks are observable, indicating a highly textured
growth. The films with thickness of 150 nm were patterned
lithographically into bridges with lateral dimensions of 33 $\mu$m
$\times$ 500 $\mu$m.

$R$-$T$ and $I$-$V$ curves were measured using a standard
four probe technique with a Keithley 182 nanovoltmeter and a
Keithley 220 current source. A pulsed current is used for the
resistance transition measurement, and the temperature of the
sample holder was stabilized to better than 0.1 K during the
measurement for each $I$-$V$ curve.

\section{Experimental results and analyses}
\begin{figure}[h!tbp]
\includegraphics[width=8cm]{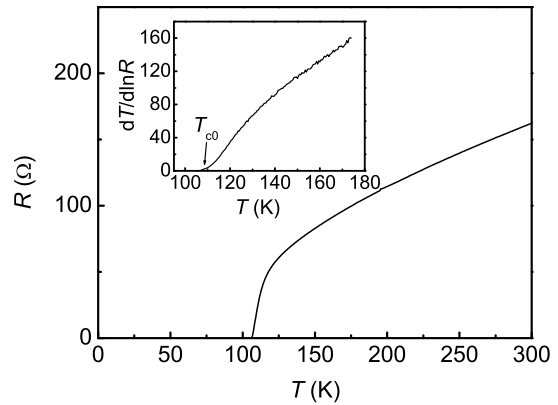}
\caption{$R$-$T$ relation of a thin Tl$_2$Ba$_2$CaCu$_2$O$_8$ film in zero
magnetic field. The inset displays $dT/d \ln R=0$ at a finite temperature, showing
a zero-resistance transition.} \label{fig:rt}
\end{figure}

The $R$-$T$ relation of the Tl$_2$Ba$_2$CaCu$_2$O$_8$ thin film in
zero magnetic field is shown in Fig.~\ref{fig:rt}. According to the
logarithmic plot in the inset, there is $d T/ d \ln R=0$ at a finite
temperature, suggesting the existence of a zero-resistance state below it.

\begin{figure}[h!tbp]
\includegraphics[width=8cm]{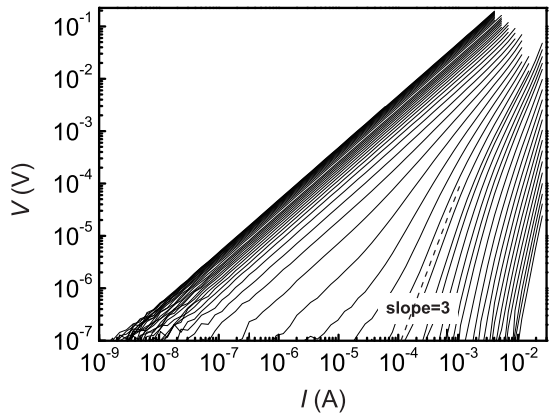}
\caption{Logarithmic $I$-$V$ isotherms of the thin
Tl$_2$Ba$_2$Ca- Cu$_2$O$_8$ film in zero magnetic field. The
broken line indicates the slope $a=3$ at which the KTB transition
occurs. The temperatures are from 95 K (the right side) to 116 K
(the left side). The increasing step of the temperatures is 0.3 K at
temperatures between 102 K and 109.5 K, otherwise it is 0.5 K.
The transition temperature $T_{\rm KT}$ determined by this
conventional approach is 103.8 K.} \label{fig:iv}
\end{figure}

The $I$-$V$ isotherms of the Tl$_2$Ba$_2$CaCu$_2$O$_8$ sample
are plotted in Fig.~\ref{fig:iv}. For our data, the determination of $a=3$
in the conventional approach is carried out at a common voltage with the
current as small as possible, which is in the vicinity of
$3\times10^{-7}$ V. The broken line in Fig.~\ref{fig:iv} shows the
condition $a=3$, therefore $T_{\rm KT}$ is determined as 103.8 K.
It is interesting that the logarithmic $I$-$V$ curves show a positive
curvature just above 103.8 K, indicating a finite linear resistivity above
$T_{\rm KT}$.

However, this conventional approach can be misleading \cite{Newrock} in
checking for $z=2$. For instance, it is difficult to determine the position of $a=3$,
and the determination of $a(T)$ for a common voltage range (instead of a
common current range) will introduce the effect of non-universal
length scales. Therefore the dynamic scaling approach proposed by
FFH \cite{Fisher} is often used, in which the restriction of
$I \rightarrow0$ does not exist and data at finite currents are also used
in the scaling. Thus we will apply this approach to our data.

\begin{figure}[h!tbp]
\includegraphics[width=8cm]{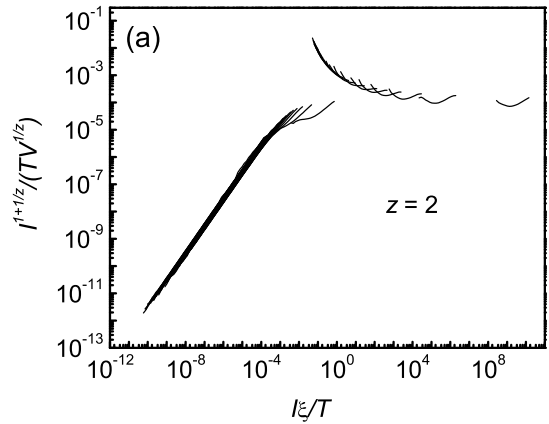}
\includegraphics[width=8cm]{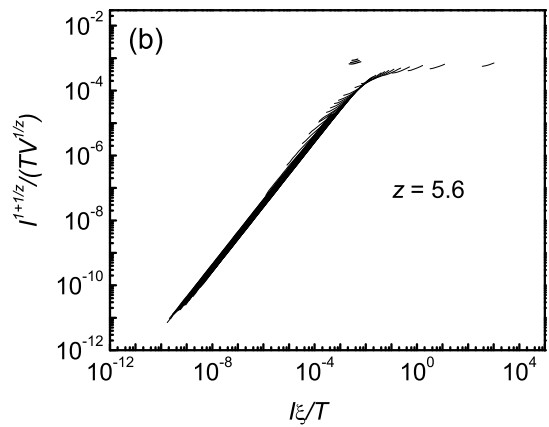}
\caption{Dynamic scaling plots with scaling
parameters (a) $z=2$ and (b) $z=5.6$. The correlation length $\xi$
is assumed to be of KTB form $\xi_{\pm}(T)\propto
\exp[\sqrt{b_{\pm}/|T/T_{\rm KT}-1|}]$, where the footnote $+(-)$
means at temperatures above (below) $T_{\rm KT}$, and $b$ is a
non-universal constant, thus $b_{+(-)}$ corresponds to the lower
(upper) branch of the plots. The other scaling parameters are:
(a) $T_{\rm KT}=103.8$ K, $b_+=0.21$, $b_-=3$; (b) $T_{\rm
KT}=96$ K, $b_+=1.2$, $b_-=0.5$ (for the upper ($T<T_{\rm KT}$)
branch the three short isotherms cannot get good collapse with any
reasonable $b_-$ value).}
\label{fig:FFH}
\end{figure}

For 2D superconductors, the dynamic scaling relation given by FFH is
\begin{equation}
V=I\xi^{-z}\chi_{\pm}(I\xi/T),\label{eq:ffh}
\end{equation}
where $\chi_{+(-)}$ is the scaling function above (below) $T_{\rm KT}$,
and $\xi$, with the original definition of correlation length in KTB theory, is
not suitable for temperatures below $T_{\rm KT}$ because of its infinite value.
It is then defined as the typical size of the vortex pairs below $T_{\rm KT}$
(for $\chi_-$)\cite{Newrock,Ambegaokar}. For convenience of scaling,
Eq.~(\ref{eq:ffh}) is often used as
\begin{equation}
\frac{I}{T}\bigg(\frac{I}{V}\bigg)^{1/z}=\epsilon_{\pm}(I \xi /
T),\label{eq:FFH}
\end{equation}
where $\epsilon_{\pm}(x) \equiv x/ \chi_{\pm}^{1/z}(x)$. To proceed
with the scaling in the form of Eq.~(\ref{eq:FFH}), $\xi$ was assumed
to be of KTB form $\xi_{\pm}(T)\propto \exp[\sqrt{b_{\pm}/|T/T_{\rm KT}-1|}]$,
where the footnote $+(-)$ means at temperatures above (below) $T_{\rm KT}$,
and $b$ is a non-universal constant. The FFH scaling results of the $I$-$V$ data
of YBCO film measured by Repaci {\it et al.} \cite{Repaci} suggest that
$z=5.6\pm0.3$ \cite{Medvedyeva}. Following this approach, we show the scaling
results of our $I$-$V$ data of Tl$_2$Ba$_2$CaCu$_2$O$_8$ film with $z=2$ and
$z=5.6$ in Figure \ref{fig:FFH}(a) and (b), respectively. The data are plotted in lines
rather than points in order to expose any shortcomings in the scaling collapse as suggested
by Holzer {\it et al.} \cite{Holzer}. The upper (lower) branch of
the plots corresponds to the temperatures below (above) $T_{\rm KT}$.

For the high temperature ($T>T_{\rm KT}$) branch, the scaling
collapse of data can be obtained with both $z=2$ and $z=5.6$. It
shows that there is a crossover from $I^{1+1/z}/(TV^{1/z})\propto
I \xi /T$ to $I^{1+1/z}/(TV^{1/z})=$ constant when $I$ increases.
This corresponds to the crossover from $V\propto I$ to $V \propto
I^{z+1}$ for each $I$-$V$ isotherm, which is shown obviously in
Fig \ref{fig:iv}, especially for those isotherms near $T_{\rm
KT}$. The crossover from linear to nonlinear $I$-$V$ relation is
reflected in both Fig.~\ref{fig:FFH}(a) and (b), and
Fig.~\ref{fig:FFH}(a) shows a much less nonlinear part than
Fig.~\ref{fig:FFH}(b).

While for the low temperature ($T<T_{\rm KT}$) branch, a much
better scaling collapse can be obtained with $z=2$ rather than
$z=5.6$. As a matter of fact, for the $T<T_{\rm KT}$ branch of
Fig.~\ref{fig:FFH}(b), a good scaling collapse is not available
for $T_{\rm KT}$ and $b_-$ with any physically possible values.
Moreover the $T_{\rm KT}$ for $z=5.6$ is much different from 103.8 K,
the value obtained from the conventional approach.

Holzer {\it et al.}\cite{Holzer} showed that for a finite-size system, the critical
exponent $z$ derived from dynamic scaling becomes larger if the noise floor is lowered.
This implies that the data with a fixed noise floor can be scaled with a $z$ value
larger than the one characterizing a true phase transition, so there is some uncertainty
if it is judged only from the dynamic scaling results Fig.~\ref{fig:FFH}(a) and (b).

Strachan {\it et al.} \cite{Strachan} argued that the FFH scaling collapse alone is
not a sufficient evidence of the phase transition, and proposed a criterion to determine the
existence of a real KTB transition. That is, the logarithmic $I$-$V$ isotherms should show
positive concavity above $T_{\rm KT}$, while show zero concavity at the same current level
below the transition temperature. Furthermore, it can be distinguished from the finite-size-caused
ohmic tail that, the current density $J_{\rm co}$, at which the $I$-$V$ relation crosses over from
ohmic to non-ohmic behavior, should increase with temperature above $T_{\rm KT}$ for a true
KTB transition, while decrease with temperature or be a constant for finite-size-caused unbinding.

\begin{figure}[h!tbp]
\includegraphics[width=8cm]{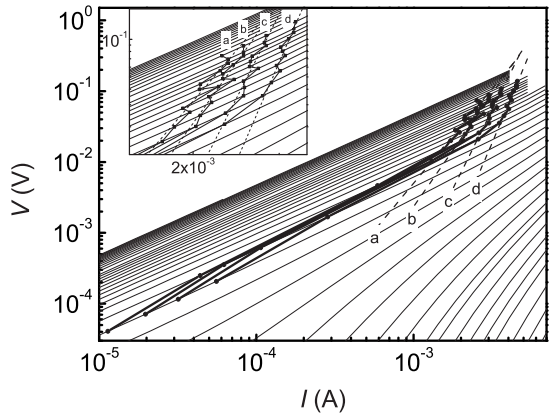}
\caption{\label{fig:non}
This Figure is used for two analysis procedures (so it suggests the similarity of the two):
I. Strachan {\it et al.}'s criterion has been used where the crossover current $I_{\rm co}$
is determined by $V/IR_{\rm lin}=$ (a) 1.015,
(b) 1.02, (c) 1.03, (d) 1.05.
II. Direct determination of $z$ from the
$I$-$V$ isotherms. The criteria are taken as $V/IR_{\rm lin}=$ (a) 1.015,
(b) 1.02, (c) 1.03, (d) 1.05, and the determined $z$ value are
(a) 1.67, (b) 2.19, (c) 3.48 and (d) 4.29, respectively. The temperatures at which the
isotherms are dotted are from 105.9 to 110 K. Below 105.9 K there is no good $R_{\rm lin}$,
and above 110 K the isotherms are too straight to reach the criteria value of $V/IR_{\rm lin}$.
There is a distinct crossover for each $V_{\rm nl}$-$I_{\rm nl}$
curve at 107.1 K, which may be the temperature that $g_L(T)$ deviates
obviously from $A_LR_{\rm lin}^{-\alpha}$ with $\alpha$ as a constant.
The inset is an enlarged view of the fitted part.}
\end{figure}

This criterion has been used in Fig.~\ref{fig:non}, where the crossover current $I_{\rm co}$
is determined by $V/IR_{\rm lin}=$ 1.015, 1.02, 1.03 or 1.05 with $R_{\rm lin}$ the linear
resistance. From 105.9 to 110 K, $I_{\rm co}$ increases with temperature. It is not available
to see the temperatures between 103.8 and 105.9 K for the absence of the linear resistance,
which may be caused by the limit of resolution.

At temperatures below 103.8 K, the logarithmic
$I$-$V$ isotherms are straight (non-ohmic) at low currents, but tending toward ohmic at high
currents, showing negative curvatures. This is similar to the $I$-$V$ simulations of Medvedyeva
{\it et al.} \cite{Medvedyeva}, which are interpreted \cite{Strachan} as a possible result of the
applicability of FFH scaling, the saturation of current-induced unbinding, or the reaching of bulk
critical current of the sample, etc.

Medvedyeva {\it et al.} \cite{Medvedyeva} also pointed out that dynamic scaling has
much flexibility and can only be used in a system with the linear size
$L \gg \xi$, so it is not compatible with a system undergoing a KTB transition because the
correlation length $\xi=\infty$ at temperatures below $T_{\rm KT}$,
and therefore only the scaling function $\chi_+$ above the
KTB transition is justified. As to the low temperature phase, the
finite size $L$ of the system serves as a cutoff in the length
scale instead of $\xi$, and a finite-size scaling form
\begin{equation}
\frac{E}{JR_{\rm lin}}=h(JLg_L(T))\label{eq:minn}
\end{equation}
has been used for smaller values of $JLg_L(T)$, where $E$, $J$ and
$L$ are the electric field, the applied current density and the
linear size of the system, respectively, $R_{\rm lin}$ is the linear resistance, and
$g_L(T)$ is a function of $T$ and $L$, which will have a finite value if $R_{\rm lin}$ is
finite, or will diverge with a vanishing $R_{\rm lin}$. To proceed with their discussion
about the finite-size scaling, $g_L(T)$ was defined as
\begin{equation}
g_L(T)=A_LR_{\rm lin}^{-\alpha_L(T)},\label{eq:gl}
\end{equation}
where $A_L$ is a universal constant which may depend on $L$. And it was claimed
that if $\alpha_L(T)$ is a $T$-independent constant, Eq.~(\ref{eq:minn}) will be
of the same form as dynamic scaling (an absolute correspondence requires
$g_L(T)=A_LR_{\rm lin}^{-\alpha_L(T)}/T$, but here the unimportant denominator $T$
was neglected), in spite of the entirely incompatible theoretical ground. Furthermore, if all the
available data are within the temperature region for which Eq.~(\ref{eq:gl}) is established, there
will be a transition at $z(T)=1/ \alpha$ which makes $g_L(T)\rightarrow \infty$ and hence
corresponds to a vanishing resistance.

Since $V=LE$ and $I=LJ$ for a 2D $L\times L$
system \cite{Medvedyeva}, Eq.~(\ref{eq:minn}) is equivalent to
\begin{equation}
\frac{V}{IR_{\rm lin}}=h(Ig_L(T)).\label{eq:minn1}
\end{equation}

As mentioned by Medvedyeva {\it et al.}, the slope $a(\equiv d \ln V/d \ln I)$
should be a constant $z(T)+1$ at small current for each isotherm below the KTB
transition. For a finite-size system, however, it will cross over to 1 in
the $I \rightarrow0$ limit instead, resulting in the maxima in the plot.
The maxima are controlled by variable $JL$ at low temperatures,
while by variable $J \xi$ at higher temperatures above $T_{\rm KT}$.
So there is also a crossover on the relevant current of each maximum, which
occurs near the temperature at which $L \approx \xi$ and is not much higher
than $T_{\rm KT}$. These two crossovers, on the value of the slope $a$ and on
the position of the maxima, have been claimed as the symbols of finite-size effect.
Moreover, by comparison of the RSJ simulation data with the conventional KTB
theory, it was concluded that the slope values $a$ at the maxima are the same
as the ones in the thermodynamic limit, so the maxima can be used to
determine the KTB transition with the conventional criterion that $z=2$
or $a=3$.

\begin{figure}[h!tbp]
\includegraphics[width=8cm]{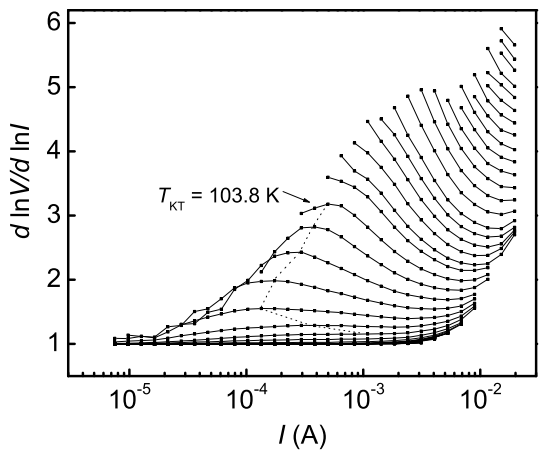}
\caption{\label{fig:peak}The $I$-$V$ data plotted as $d \ln V/d \ln I$
against $I$ in logarithmic scale. The relevant temperatures
decrease from the lower curves to higher ones. The broken line
segments are used to connect the maxima which give estimates of
$z(T)+1$. $T_{\rm KT}$ is determined by the temperature at which the
maximum of the slope $d \ln V/d \ln I$ is 3. The result is consistent
with the $T_{\rm KT}$ value got by the conventional approach that
$T_{\rm KT}= 103.8$ K. The crossover of the maxima occurs at 105
K, a temperature slightly higher than $T_{\rm KT}$.}
\end{figure}

Fig.~\ref{fig:peak} shows the slope $a \equiv d \ln V/d \ln I$ got from Fig.~\ref{fig:iv}
versus $I$ in logarithmic scale. The data beyond the noise background have been
removed. $T_{\rm KT}$ is 103.8 K, determined by the temperature of the $a$-$I$
curve which has the maximum of $a(=z+1)=3$. The result is consistent with that of
previous approaches. Unlike the isotherms above $T_{\rm KT}$, there are no peaks
appeared on the isotherms below $T_{\rm KT}$. When the temperature increases,
the current at which the maxima appear decreases slightly at first,
then increases rapidly (taking note of the logarithmic scale of $I$ in the plot),
and the crossover occurs at the temperature of 105 K, which is
somewhat higher than $T_{\rm KT}$. This agrees with the above
picture of crossover on length scale ($L$-dominant to $\xi$-dominant)
mentioned by Medvedyeva {\it et al.}~from both simulation and the experimental
results. A more detailed comparison with the work of Medvedyeva {\it et al.}~suggests
that in the $L$-dominant part of $a$-$I$ plot for Repaci's YBCO
data, the maxima appear at almost the same current, while for the 2D
RSJ model, the maxima move toward the low current direction slightly
with increasing temperature. Our data are more similar to the
latter with more obvious decreasing.

The upturn of the curves at large currents may be caused by the
heat effect hence it is not taken into account during the
analyses.

\begin{figure}[h!tbp]
\includegraphics[width=8cm]{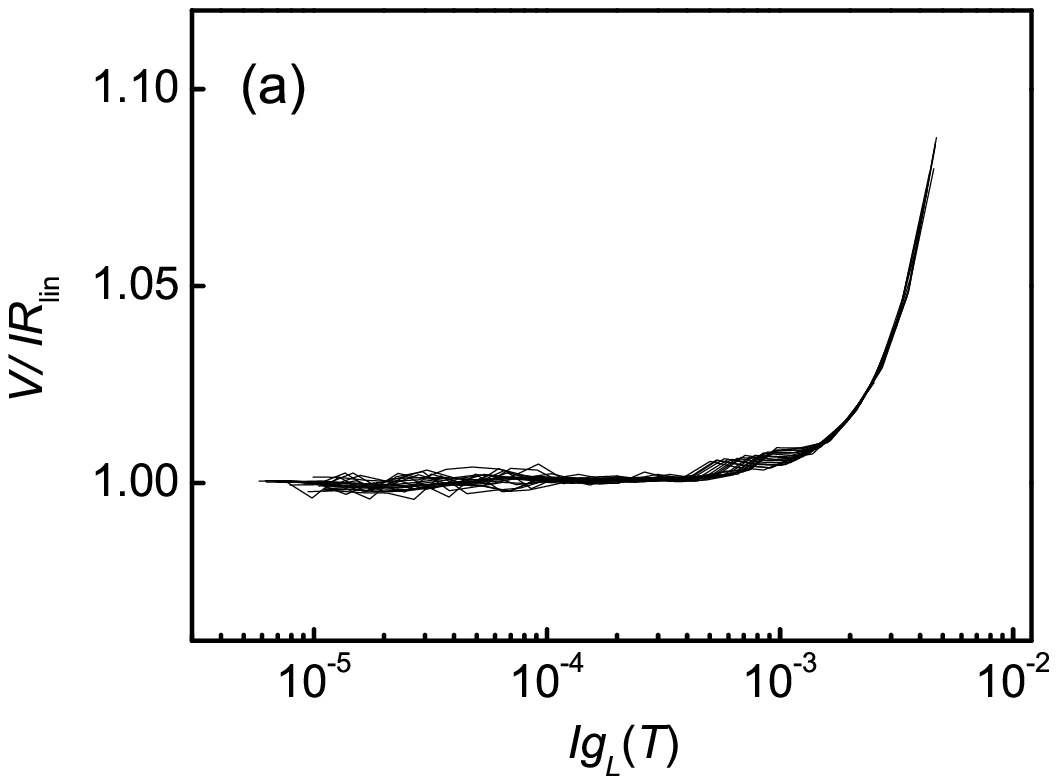}
\includegraphics[width=8cm]{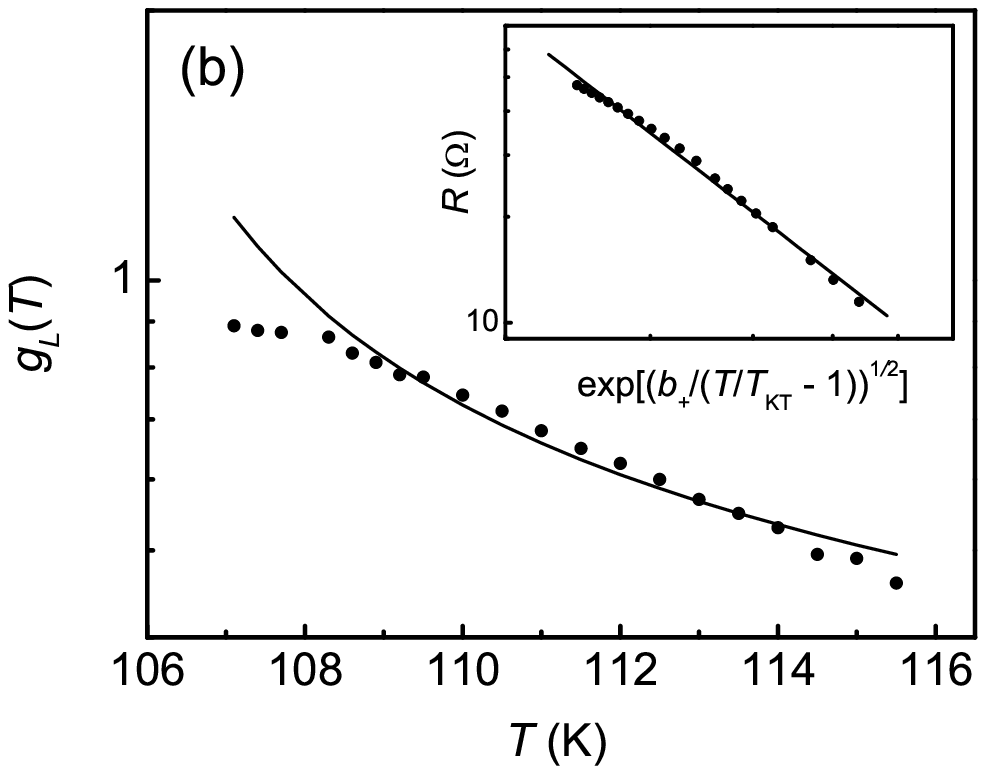}
\caption{\label{fig:gl}(a) Determination of the
existence of the scaling in the form of Eq.~\ref{eq:minn1} for our
$I$-$V$ data. $V/IR_{\rm lin}$ is plotted against $Ig_L(T)$, and $g_L(T)$ is
derived from the collapse of data. The relevant temperatures are
from 107.1 to 115.5 K, which corresponds to the $\xi < L$ regime
above $T_{\rm KT}$. (b) The function $g_L(T)$ plotted versus $T$.
The filled circles give the values of $g_L(T)$ determined from scaling collapse
of $I$-$V$ data in (a). The fitting curve shows that
$g_L(T)$ can be represented by $A'_L \exp(\sqrt{b_+/(T/T_{\rm KT}-1)})/T$
with $A'_L=23.4$ and $b_+=0.09$. The inset shows the linear resistance
$R_{\rm lin}$ versus $\exp(\sqrt{b_+/(T/T_{\rm KT}-1)})$ plotted in $\log_{10}$-$\log_{10}$
scale. The slope of the fitting curve is $-1.8$, which suggest that $\alpha=1/1.8$
and $z=1.8$, which is much closer to the result $z=2$ rather than $z=5.6$.}
\end{figure}

Subsequently the question is whether the scaling given by
Eq.~(\ref{eq:minn1}) works. Fig.~\ref{fig:gl}(a) shows the scaling collapse
for our $I$-$V$ data. $V/IR_{\rm lin}$ is plotted against $Ig_L(T)$ and $g_L(T)$ is
derived from the collapse of data. Because the linear resistance $R_{\rm lin} \equiv
\lim_{I \rightarrow0}d \ln V/ d \ln I$ is required
in the approach, which amounts to a requirement of a small current,
the scaling is proceeded with the isotherms which can
derive good linear resistance. The relevant temperatures are from
107.1 to 115.5 K which are above $T_{\rm KT}$ and in the regime
where $\xi < L$. Nevertheless, the data show a good collapse for the finite-size
scaling approach from a pragmatic point of view \cite{Medvedyeva}.
This connection of finite-size scaling with dynamic scaling implies the requirement
that $\alpha_L(T)$ in Eq.~(\ref{eq:gl}) being a constant $\alpha$ is met.

For the temperatures lower than 107.1 K, including the region
where the sample shows finite-size effect (at 105 K or
temperatures lower, according to Fig.~\ref{fig:peak}), the scaling
collapses have not been obtained. This may be induced by the fact
that the linear resistance $R_{\rm lin}$ and the voltage data at
low enough currents, which are important elements in the
finite-size scaling equation $V/IR_{\rm lin}=h(Ig_L(T))$, are not
available.

To testify that $\alpha_L(T)$ is a constant and derive its value, the function of the form
$A_LR_{\rm lin}^{-\alpha}/T$ should be used to fit $g_L(T)$ (in Eq.~(\ref{eq:gl}) Medvedyeva
{\it et al.}~neglected the unimportant denominator $T$ in $g_L(T)$, but here we keep
it for better accuracy in fitting), where $R_{\rm lin}$ is the linear resistance
determined from small current limit of the $I$-$V$ isotherms, $A_L$ and $\alpha$ are
two fitting parameters. For convenience of fitting with smooth functions, we divide
it into two steps. At first, the function $A'_L \exp(\sqrt{b_+/(T/T_{\rm KT}-1)})/T$
is used for fitting, with $A'_L$ and $b_+$ as two fitting parameters and regardless of the
value of $\alpha$; secondly, linear fit the $\log R_{\rm lin}$-$\log[\exp(\sqrt{b_+/(T/T_{\rm KT}-1)})]$
relation and derive the slope, which corresponds to $-1/ \alpha$. The above procedures come from
the consideration that $R_{\rm lin} \propto \xi^{-z}$, $\xi \propto \exp(\sqrt{b_+/(T/T_{\rm KT}-1)})$
and $z=1/ \alpha$ in the context of Medvedyeva {\it et al.}\cite{Medvedyeva}. These
two steps are shown in Fig.~\ref{fig:gl}(b). The fitness shows that
$g_L(T)$ is most likely to be represented by $\alpha=1/1.8$ ($z=1.8$), which is much
closer to $z=2$ rather than $z=5.6$. At lower temperatures there is a visible
deviation from the fit, which may be caused by finite-size effect.

According to Medvedyeva {\it et al.}\cite{Medvedyeva}, the above results
suggest that, if going down from a high temperature, a resistive transition will
occur at some lower $T$ outside the region where Eq.~(\ref{eq:gl}) holds
(with $\alpha=1/1.8$). That should correspond to the KTB transition ($z=2$).

Because the $I$-$V$ data can be scaled with Eq.~(\ref{eq:minn1}), the linear
part at small current corresponds to $V/IR_{\rm lin}=1$. And if $V/IR_{\rm lin}=c$ is taken as the
criterion of the onset of the nonlinear relation on the $I$-$V$ isotherms, where
$c$ is a constant, we will have
\begin{equation}
\frac{V_{\rm nl}}{I_{\rm nl}R_{\rm lin}}=h(I_{\rm nl}A_LR_{\rm lin}^{-\alpha_L(T)})=c,
\end{equation}
where $h$ is the scaling function defined in Eq.~(\ref{eq:minn1}), and ($I_{\rm nl}$,
$V_{\rm nl}$) denotes the onset point on the $I$-$V$ isotherm, thus we have
\begin{equation}
\ln V_{\rm nl}=\Big[1+\frac{1}{\alpha_L(T)}\Big]\ln I_{\rm nl}+\Big[\ln c-\frac{1}
{\alpha_L(T)}\ln \frac{c'}{A_L}\Big],
\end{equation}
where $c'$ is the constant which satisfies $h(c')=c$. Therefore if $\alpha$ is
a constant within a temperature region, the logarithmic $V_{\rm nl}$-$I_{\rm nl}$
curve should be a straight line with the slope$=(1+1/ \alpha)$ at these temperatures.
So $\alpha$ can be determined directly from the slope of $V_{\rm nl}$-$I_{\rm nl}$
curve. This is exactly what Medvedyeva {\it et al.} did in Fig.~13 of Ref.~\onlinecite{Medvedyeva}.

Fig.~\ref{fig:non} shows the $V_{\rm nl}$-$I_{\rm nl}$ curves with various criteria that
$V/IR_{\rm lin}=c=1.015, 1.02, 1.03, 1.05$, respectively. The relevant temperatures are from 105.9
to 110 K. There is a distinct crossover for each $V_{\rm nl}$-$I_{\rm nl}$ curve at 107.1 K, which
is approximately the temperature that $g_L(T)$ begins to deviate obviously from
$A_LR_{\rm lin}^{-\alpha}$ (see Fig.~\ref{fig:gl}(b)) with a constant $\alpha$ value. As mentioned
above, direct determination of $\alpha$ (or $z$) from the slope of $V_{\rm nl}$-$I_{\rm nl}$
curves only makes sense when $\alpha$ is a constant, thus the linear fit of
$V_{\rm nl}$-$I_{\rm nl}$ curves should be carried out above 107.1 K, as shown by the
dash lines. The results should be more reliable as the parameter $c$ decreases
because the scaling function Eq.~(\ref{eq:minn1}) is satisfied at currents not very high
\cite{Medvedyeva}. And the small $c$ results show that $z \approx 2$.

\begin{figure}[h!tbp]
\includegraphics[width=8cm]{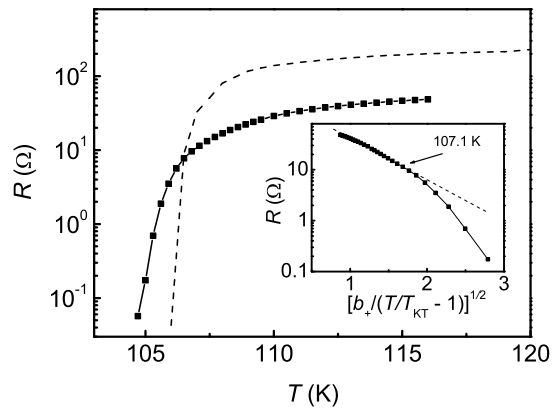}
\caption{\label{fig:rlin}Temperature dependence of the linear resistance. The dotted line
is derived from the low current limit of Fig.~\ref{fig:iv}, with the relevant temperatures
from 104.7 to 116 K. The dashed line is from $R$-$T$ measurement taken at a finite current which
is plotted in Fig.~\ref{fig:rt}.
The inset shows the logarithmic linear resistance $R_{\rm lin}$ versus $\sqrt{b_+/(T/T_{\rm KT}-1)}$,
with a wider temperature range (105 to 116 K) than that of the inset of Fig.~\ref{fig:gl}(b). }
\end{figure}

Fig.~\ref{fig:rlin} shows the temperature dependence of linear resistance. The data obey
Kosterlitz relation ($R_{\rm lin}\propto \exp(\sqrt{b_+/(T/T_{\rm KT}-1)})$) above 107.1 K,
but deviate from the linear relation below 107.1 K. This may be interpreted as the crossover to
finite-size effect. The directly measured $R$-$T$ data with a finite current drop more sharply than
$R_{\rm lin}$ do.

\begin{figure}[h!tbp]
\includegraphics[width=8cm]{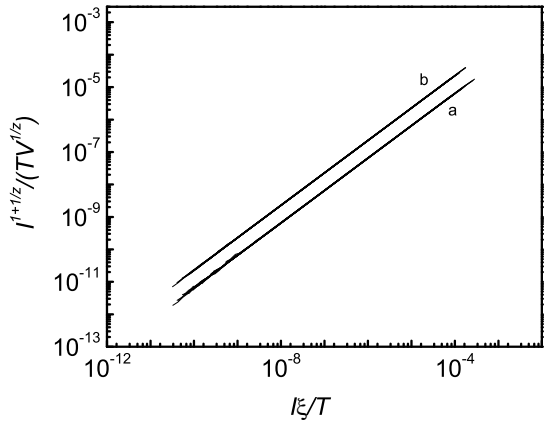}
\caption{\label{fig:FFH-better}Re-scaling of $I$-$V$ data
with FFH method at higher temperatures where
$\xi \ll L$ and all the data influenced by the finite-size effect
are removed, including the data at temperatures slightly above
that corresponding to $\xi \sim L$ (about 105 K), which still have
some effect of finite size and are not collapsed well. The temperatures
of the shown data are from 107.1 to 116 K, and the scaling
parameters are:
(a) $T_{\rm KT}=103.8$ K, $b_+=0.07$, $z=2$;
(b) $T_{\rm KT}=96$ K, $b_+=0.12$, $z=5.6$.}
\end{figure}

Since the above analyses suggest that the finite-size effect takes place in the $I$-$V$
data, the scaling collapse in Fig.~\ref{fig:FFH} which is not so perfect can be attributed
to that the data showing finite-size effect were scaled together with the data in
the thermal dynamic limit by dynamic scaling method. So for the data at temperatures above
$T_{\rm KT}$, if we remove those data showing finite-size effect and only consider the
data at higher temperatures where $\xi_+ \ll L$, the scaling collapse should be
better. This is exhibited in Fig.~\ref{fig:FFH-better} with the relevant scaling parameter
$b_{+}$ changed. It shows that both $z=2$ and $z=5.6$ can give good collapse, which also
suggest that some criterion is needed to eliminate the uncertainty.

A pure KTB transition, in the thermodynamic limit, requires \cite{Newrock} $L \to \infty$ and
$\lambda_\perp \gg L$. For the TBCCO sample with intrinsically high anisotropy, the estimated value of
$\lambda_\perp(\sim \lambda^2/d)$ is about 60 $\mu$m with $\lambda=1500$ \AA \ and $d=3.8$ \AA,
and the lateral sample size $L=33$ $\mu$m, thus the sample size $L$ is not much larger compared with
the size of vortex pairs and  $\lambda_\perp$ is not much greater than $L$, especially at zero temperature.
Therefore the evidence of finite-size effect can be attributed to the finite cut-off length scale, which is $L$
rather than $\lambda_\perp$ for our data near the critical temperature.

\section{Concluding Remarks}

We have performed the conventional approach, FFH dynamic scaling
approach and finite-size scaling approach proposed by Medvedyeva {\it et al.}
on the data of thin Tl$_2$Ba$_2$CaCu$_2$O$_8$ films in zero magnetic field.
The data in the small current limit together with those of finite values
are studied. The results from all these methods are consistent with each
other that the $I$-$V$ characteristics are dominated by the vortex-unbinding
with the nominal transition temperature $T_{\rm KT}=103.8$ K, and the dynamic scaling
exponent $z=2$ hence the $I$-$V$ exponent $a=3$ ($a \equiv d \ln V/d \ln I$).

Besides, the evidence of finite-size effect is also found, that the maxima of slope
$a$ of logarithmic $I$-$V$ isotherms cross over from $L$-dependent to $\xi$-dependent
with the crossover temperature at 105 K, slightly above $T_{\rm KT}$. The finite-size-caused
free vortices will lead to the absence of a phase transition in a strict sense, and will turn
the phase transition into a crossover behavior \cite{Newrock}. However,
the finite-size effect in our experiments is not so remarkable that, even at
the voltage measuring limit the $I$-$V$ isotherms are not showing a linear
resistivity. Medvedyeva {\it et al.} \cite{Medvedyeva} gave an explanation for
the $I$-$V$ data of Repaci {\it et al.}~which show $z \approx6$. The
transition has been called a ``ghost'' transition because it is only a finite-size
effect and does not occur in the thermodynamic ($L \gg \xi$) limit. In our
case, however, the transition at the temperature below $\alpha=1/1.8$ expected
by the finite-size scaling should correspond to the vortex-unbinding transition with $z=2$, behaving
more likely as the thermodynamic limit rather than a ``ghost'' transition. This can be interpreted
by the high experimental temperature near $T_{\rm c}$ so that the size of our sample is not in
the finite-size limit as that in the simulation work of Medvedyeva {\it et al.}, therefore it is sufficient
to bring a finite-size effect in the KTB transition but does not markedly change the $z=2$ result.

\begin{acknowledgments}
This work is financially supported by the National
Science Foundation of China (NSFC 19825111) and the Ministry of
Science and Technology of China (project: NKBRSF-G1999064602).
We acknowledge fruitful discussion with J.~M.~Kosterlitz, Brown
University.
\end{acknowledgments}


\begin{thebibliography}{99}

\bibitem{Mermin}
 N. D. Mermin,
 Phys. Rev. {\bf176}, 250 (1968).
 N. D. Mermin and H. Wagner,
 Phys. Rev. Lett. {\bf 22}, 1133 (1966).
 P. C. Hohenberg, Phys. Rev. {\bf158}, 383 (1967).

\bibitem{Berezinskii}
 V. L. Berezinskii,
 Sov. Phys. JETP {\bf 34}, 610 (1972).

\bibitem{Kosterlitz}
 J. M. Kosterlitz and D. J. Thouless,
 J. Phys. C {\bf 6}, 1181 (1973).

\bibitem{Kosterlitz1}
 J. M. Kosterlitz and D. R. Nelson,
 Phys. Rev. Lett. {\bf 39}, 1201 (1977).

\bibitem{Pearl}
 J. Pearl,
 Appl. Phys. Lett. {\bf 5}, 65 (1964).

\bibitem{Halperin}
 B. I. Halperin and D. R. Nelson,
 J. Low Temp. Phys. {\bf 36}, 599 (1979).

\bibitem{Hebard}
 A. F. Hebard and A. T. Fiory,
 Phys. Rev. Lett. {\bf 50}, 1603 (1983).
 P. A. Bancel and K. E. Gray,
 Phys. Rev. Lett. {\bf 46}, 148 (1981).

\bibitem{Rasolt}
 M. Rasolt, T. Edis, and Z. Tesanovic,
 Phys. Rev. Lett. {\bf 66}, 2927 (1991).

\bibitem{Emery}
 V. J. Emery and S. A. Kivelson,
 Nature {\bf 374}, 434 (1995).

\bibitem{Corson}
 J. Corson, R. Mallozzi, J. Orenstein,
 J. N. Eckstein, and I. Bozovic,
 Nature {\bf 398}, 221 (1999).

\bibitem{Xu}
 Z. A. Xu, N. P. Ong, Y. Wang, T. Kakeshita, and S. Uchida,
 Nature {\bf 406}, 486 (2000).

\bibitem{Wang}
 Y. Wang, Z. A. Xu, T. Kakeshita, S. Uchida, S. Ono, Yoichi
 Ando, and N. P. Ong,
 Phys. Rev. B {\bf 64}, 224519 (2001).

\bibitem{Millis}
 A. J. Millis,
 Nature {\bf 398}, 193 (1999).

\bibitem{Minnhagen}
 P. Minnhagen,
 Rev. Mod. Phys. {\bf 59}, 1001 (1987).

\bibitem{Fisher}
 D. S. Fisher, M. P. A. Fisher, and D. A. Huse,
 Phys. Rev. B {\bf 43}, 130 (1991).

\bibitem{Matsuda}
 Y. Matsuda and S. Komiyama,
 Phys. Rev. B {\bf48}, 10498 (1993).

\bibitem{Martin}
 S. Martin, A. T. Fiory, R. M. Fleming, G. P. Espinosa,
 and A. S. Cooper,
 Phys. Rev. Lett. {\bf62}, 677 (1989).

\bibitem{Kim}
 D. H. Kim, A. M. Goldman, J. H. Kang, and
 R. T. Kampwirth,
 Phys. Rev. B {\bf40}, 8834 (1989).

\bibitem{Wen}
 H. H. Wen, P. Ziemann, H. A. Radovan and S. L. Yan,
 Europhys. Lett. {\bf42}(3), 319 (1998).

\bibitem{Repaci}
 J. M. Repaci, C. Kwon, Q. Li, X. Jiang, T. Venkatessan,
 R. E. Glover, C. J. Lobb, and R. S. Newrock,
 Phys. Rev. B {\bf 54}, R9674 (1996).

\bibitem{Ammirata}
 S. M. Ammirata, M. Friesen, S. W. Pierson,
 LeRoy A. Gorham, Jeffrey C. Hunnicutt,
 M. L. Trawick, and C. D. Keener,
 Physica C {\bf 313}, 225 (1999).

\bibitem{Holzer}
 J. Holzer, R. S. Newrock, C. J. Lobb, T. Aouaroun, and
 S. T. Herbert,
 Phys. Rev. B {\bf63}, 184508 (2001).

\bibitem{Medvedyeva}
 K. Medvedyeva, B. J. Kim, and P. Minnhagen,
 Phys. Rev. B {\bf 62}, 14531 (2000).

\bibitem{Strachan}
 D. R. Strachan, C. J. Lobb, and R. S. Newrock,
 unpublished.

\bibitem{Yan}
 S. L. Yan, L. Fang, Q. X. Song, J. Yan, Y. P. Zhu, J. H. Chen,
 and S. B. Zhang,
 Appl. Phys. Lett. {\bf 63}, 1845 (1993).

\bibitem{Newrock}
 R. S. Newrock, C. J. Lobb, U. Geigenm$\rm \ddot{u}$ller, and M. Octavio,
 Solid Stat. Phys. {\bf54}, 263 (2000).

\bibitem{Ambegaokar}
 V. U. Ambegaokar, B. I. Halperin, D. R. Nelson, and E. D. Siggia,
 Phys. Rev. Lett. {\bf40}, 783 (1978).
 V. U. Ambegaokar, B. I. Halperin, D. R. Nelson, and E. D. Siggia,
 Phys. Rev. B {\bf21}, 1806 (1980).

\end{thebibliography}
\end{document}